\documentclass[a4paper,11pt]{article}
\pdfoutput=1
\usepackage{jheppub}
\usepackage[T1]{fontenc}
\usepackage{amsfonts,amsmath,amssymb}
\usepackage{multirow}
\usepackage{graphicx}
\usepackage{mathtools}
\usepackage{comment}
\usepackage{subcaption}
\usepackage{xcolor}

\renewcommand{\O}{{\cal O}}

\def\be{\begin{equation}}
\def\ee{\end{equation}}
\def\bea{\begin{eqnarray}}
\def\eea{\end{eqnarray}}

\numberwithin{equation}{section}

\preprint{LA-UR-25-31761}

\begin{document}

\title{Quantum bootstrap for central potentials}

\author[a]{Scott Lawrence,}
\emailAdd{srlawrence@lanl.gov}
\affiliation[a]{Theoretical Division, Los Alamos National Laboratory, Los Alamos, NM 87545}
\author[b]{Brian McPeak}
\emailAdd{bmmcpeak@syr.edu}
\affiliation[b]{Syracuse University, Syracuse, NY 13244, USA}

\date{\today}

\abstract{
We study the quantum-mechanical bootstrap as it applies to the bound states of several central potentials in three dimensions. As part of this effort, we show how the bootstrap approach may be applied to ``non-algebraic'' potentials, such as the Yukawa potential (which asymptotically decays as an exponential) and a Gaussian potential. We additionally review the bootstrap of the Coulomb potential, demonstrate a high-precision bootstrap of the Cornell potential, and study conformal quantum mechanics. These results further recommend the bootstrap as a numerical method for high-precision calculations of ground-state physics, where applicable: for example, we are able to determine the critical coupling in the Cornell potential to better than one part in $10^7$, the most precise determination to date. Lower bounds on energies are also of high precision, occasionally one part in greater than $10^8$. Finally, we discuss the circumstances under which we are able to obtain meaningful upper bounds on ground-state energies.
}
\maketitle

\section{Introduction}

The ground state of a quantum-mechanical system may be characterized by a set of expectation values, which must obey certain constraints. Some expectation values are related to each other by commutation relations; others are related by the Heisenberg relation $\langle [\hat H, \hat {\mathcal O}]\rangle = 0$; and some are constrained by positivity of the norm on the Hilbert space, which implies that $\langle \mathcal O^\dagger \mathcal O\rangle \ge 0$. Considering a basis of $N$ different Hermitian operators $\hat B_1,\ldots,\hat B_N$, there are $N$ different expectation values which define a space isomorphic to $\mathbb R^N$. Not all points in this space satisfy the above constraint; generically, the locus which is consistent with the above constraints is some lower-dimensional convex subspace of $\mathbb R^N$.

Taken together, these constraints provide the basis for a computational method known as the (quantum-mechanical) bootstrap. The bootstrap is not an analytical technique which provides exact answers; neither is it a Monte Carlo method that provides an unbiased statistical estimate. Rather, the bootstrap is a technique for obtaining rigorous \emph{bounds} on expectation values in a quantum system, and in its most common incarnation, bounds on the ground-state energy. Closely related to the conformal bootstrap~\cite{Kos:2016ysd}, the quantum-mechanical bootstrap has been used to study a wide variety of quantum systems, including one-dimensional quantum mechanics~\cite{Berenstein:2021dyf,Berenstein:2021loy}, matrix models~\cite{Han:2020bkb, Li:2025tub}, integrable systems \cite{Aikawa:2025dvt}, other theories of matrix quantum mechanics~\cite{Lin:2023owt,Lin:2024vvg, Lin:2025srf, Laliberte:2025xvk}, scattering \cite{Berenstein:2023ppj}, open systems \cite{Robichon:2024apx}, lattice systems~\cite{PhysRevLett.108.200404,Anderson:2016rcw,Lawrence:2021msm,Lawrence:2022vsb,Kazakov:2022xuh,Cho:2022lcj, Cho:2023ulr, Kazakov:2024ool, Li:2024wrd,Guo:2025fii}, systems at finite temperature~\cite{Fawzi:2023fpg,Cho:2024kxn} and with many degrees of freedom \cite{Kull:2022wof,Wang:2023hss, Cho:2024owx, Scheer:2024eyu,Berenstein:2024ebf,Jansen:2025nlc, Gao2025}, and real-time and out-of-equilibrium dynamics~\cite{Lawrence:2024hjm,Lawrence:2024mnj,Cho:2025dgc,Cho:2025nlv}. For a recent pedagogical introduction to these methods, see~\cite{Lin:2025iir}.

Despite the wide applicability of bootstrap-like techniques, fundamental questions about the behavior of these algorithms remain. In particular: as larger bases of operators are examined, do the bounds obtained by bootstrap methods become arbitrarily tight? Or will the allowed regions converge to islands of finite size? In principle the answer may be system dependent, and there may exist particularly unpleasant systems for which the bootstrap provides no bounds whatsoever.

The purpose of this paper is to shed some light on the situation by expanding the set of potentials for which the bootstrap can be seen to obtain qualitatively tight bounds. In no case do we present a proof that these bounds can be made arbitrarily tight, but we are able to numerically show that these bounds can be improved to better than one part in a million across a wide range of coupling parameters.

To make these questions more concrete, we now recount the numerical method at the core of a typical bootstrap calculation in detail. We begin by selecting a \emph{generating set} of $K$ operators $A_1,\ldots,A_K$. We will consider the $K^2$ expectation values of the form $\langle A^\dagger_i A_j\rangle$, and the various relations between, and constraint on, those expectation values.

Certain sets of these expectation values are related by affine equalities. Consider the case of the anharmonic oscillator, where we work with operators which are polynomials of $\hat x$ and $\hat p$. Then the expectation values $\langle \hat x \hat p\rangle$ and $\langle \hat p \hat x\rangle$ are not linearly independent, but are instead constrained by the canonical commutation relation:
\begin{equation}
    \langle \hat x \hat p\rangle - \langle \hat p \hat x\rangle = i\text.
\end{equation}
Similarly, the Heisenberg equations of motion provide Hamiltonian-dependent linear constraints on expectation values in any eigenstate\footnote{In fact these constraints hold true for any density matrix which commutes with the Hamiltonian.}. All these constraints are of the form $\langle [\hat H,\hat O]\rangle = 0$.

Finally, these expectation values are constrained by positivity of the Hilbert-space norm. For any operator $\mathcal O$, the expectation value $\langle \mathcal O^\dagger \mathcal O\rangle$ must be non-negative in any state. It follows that the matrix of expectation values constructed from the basis $\{A_i\}$ must be positive semi-definite:
\begin{equation}
    M_{ij} \equiv \langle A^\dagger_i A_j \rangle \succeq 0
\end{equation}

In the ground state of a Hamiltonian $\hat H$, it may also be shown~\cite{Fawzi:2023fpg} that another matrix is positive semi-definite:
\begin{equation}
    G_{ij} \equiv \langle A_i^\dagger [\hat H,A_j]\rangle \succeq 0
\end{equation}
If the generating set of operators covers the full Hilbert space, then this latter statement is true \emph{only} for the ground state.

These constraints---commutation relations, the equations of motion, Hilbert-space positivity, and positivity of $G$---together define a convex space. A bootstrap calculation proceeds by exploring this convex space\footnote{Or its dual, in the sense of Lagrange duality. This is not essential for the examples in this work, but is necessary in order to rigorously bound infinite-dimensional spaces, such as those involved in the time-evolution of quantum systems~\cite{Lawrence:2024mnj}.} to discover the minimum (and maximum) attainable value of some expectation value $\langle \mathcal O \rangle$. This can be done efficiently by means of one of many convex optimization algorithms~\cite{boyd2004convex}; in this work we use the SDPB software package~\cite{Simmons-Duffin:2015qma} for this task.

It is a straightforward matter to prove that, when the set $\{A_i\}$ forms a complete basis fo the space of operators on Hilbert space, the constraints above uniquely identify the ground state. In this case, the upper and lower bounds on the expectation of an arbitrary operator $\mathcal O$ will be equal, and equal to the true value that $\langle \mathcal O \rangle$ takes in the ground state.

For many systems of interest, no finite set of operators forms a complete basis, precisely because the Hilbert space is infinite-dimensional. The simplest such systems describe the quantum mechanics of a single particle in a small number of spatial dimensions, which is the focus of this work, but the phenomenon includes the infinite-volume limits of spin systems, as well as relativistic quantum field theories. For these systems we may now ask the following question. Consider a sequence of bases $\mathcal A^{(k)} \equiv \{A_i^{(k)}\}$, indexed by $k$, such that each basis is a strict subset of the next: $\mathcal A^{(k)} \subset \mathcal A^{(k+1)}$. For any observable $\mathcal O$, the basis $\mathcal A^{(k)}$ defines upper and lower bounds on the corresponding expectation value, which we denote $\mathcal O_<$ and $\mathcal O_>$. Under what circumstances do these bounds converge; that is, when can we guarantee that
\begin{equation}
\lim_{k \rightarrow \infty} \left[\mathcal O_< - \mathcal O_>\right] = 0
    \text?
\end{equation}

In this paper we investigate this question by example, and with a particular eye towards expanding the set of systems for which bases can be constructed which appear (from numerical evidence) to give convergent bounds. In Section~\ref{sec:coulomb} we review the bootstrap of the Coulomb potential, which can be done either by reduction to the radial Schr\"odinger equation or by considering operators defined in three-dimensional Cartesian coordinates. In both cases tight bounds are available with a fixed finite basis. 

Next we move to potentials which are not exactly solvable, and have more exotic structure. In all cases we work with the radial Schr\"odinger equation, and care is needed to deal with the boundary conditions at the origin. On the relevant Hilbert space, the momentum operator canonically conjugate to the radial coordinate is \emph{not} self-adjoint, due to boundary terms picked up at the origin. This phenomenon has been termed an anomaly~\cite{Berenstein:2022ygg}, and its correct treatment is necessary in order to obtain correct bounds on the ground state.

Section~\ref{sec:yukawa} considers the Yukawa potential, which frequently occurs as a low-energy effective potential when a massive force-carrying boson has been integrated out. We are able to achieve qualitatively tight (lower) bounds with a relatively small basis. Nuclear interactions are often modeled by a Gaussian potential, which we study in Section~\ref{sec:gaussian}, again achieving tight bounds on the ground state energy. A bootstrap calculation of the Cornell potential---a good model for the spectrum of heavy mesons like charmonium---is presented in Section~\ref{sec:cornell}. In this case we obtain both upper and lower bounds on the ground-state energy, making use of the additional constraints suggested in~\cite{Fawzi:2023fpg}. Our last example is conformal quantum mechanics. Defined by a potential proportional to $\frac 1 {r^2}$, this system is notable for having a spectrum which is unbounded below until a special boundary condition at $r=0$ is selected. Finally in Section~\ref{sec:discussion} we review these results, and consider their implications for the convergence properties of bootstrap calculations.

\section{Review of the Coulomb potential}\label{sec:coulomb}

We begin by reviewing two perspectives on the bootstrap of a single spin-0 particle in a Coulomb potential. The Hamiltonian of the particle is
\begin{equation}
    \hat H_{\mathrm{Coulomb}}
    = \sum_i \frac{\hat p_i^2}{2M}
    - \frac{\alpha}{\hat r}
    \text,
\end{equation}
where $M$ is the mass of the particle, $\alpha$ is the fine structure constant, and the index $i$ ranges over the three spatial dimensions.

In the first subsection below we perform the standard reduction to spherical coordinates, and bootstrap only the $l=0$ sector. In the next subsection, we perform the same calculation in three dimensions, without the restriction to the $l=0$ sector, and discuss the relation between the two calculations.

\subsection{Radial coordinates}

We begin by reducing the Coulomb Hamiltonian to its one-dimensional equivalent, in the $l=0$ sector. The wavefunction $\psi$ in the $l=0$ sector will be rotationally invariant, so that a single function $\psi(r)$ of the radial coordinate suffices. It is conventional to rescale this wavefunction by $u(r) = r \psi(r)$. Under this rescaling, the Hamiltonian acting on $u$ is
\begin{equation}
    \hat H_{\mathrm{Coulomb},l=0}
    = \frac{\hat p^2}{2 M} - \frac\alpha{\hat r}
    \text,
\end{equation}
where $\hat p$ is the momentum conjugate to $r$: $p = -i \frac{\partial}{\partial r}$.

Bootstrap bounds for this system have previously been derived~\cite{Berenstein:2021dyf, Berenstein:2022ygg} but we will write things in a slightly different way, with an eye towards the cases we will explore later in this work. In particular, the basis of functions we will consider is 
\begin{align}
    \left\{ \mathcal O \right \} \ = \ \left\{ \hat p^a \, , \hat r^b \, , \hat{V}^{(c)} \right \} \, ,
\end{align}
where $V^{(c)}(r)$ is the $c^{th}$ derivative of $V$. The two usual sources of linear constraints for this type of system are 
\begin{align}
    \label{eq:reality} 
    \text{reality:}& \qquad \langle \mathcal O \rangle \ = \ \langle \mathcal O^\dagger \rangle^* \\
    \label{eq:eom}
    \text{equations of motion:}& \qquad  \langle [H ,\mathcal O] \rangle \ = \ 0 \, .
\end{align}
The basis of functions we describe above has the pleasant property that these constraints will be agnostic about the particular potential under consideration. Theory-dependent information then enters into the form of the differential equation satisfied by the potential. In the present case, we have 
\begin{align}
    \hat r \hat V^{(n+1)} \ = \ -n \hat V^{(n)} \, .
\end{align}
This statement is true as an operator equality (\textit{i.e.}~not only inside expectation values) and it can be used to reduce the number of independent variables significantly.

A technical issue present for Coulomb and many other central potentials is that $V(r)$ diverges at the origin. This means that the wave-functions must go to zero at the origin fast enough that the energy remains finite. For the Coulomb potential, the requirement is that 
\begin{align}
    u(r) \sim r^a \, , \qquad a > \frac{1}{2} \, .
\end{align}
Indeed one can see that the actual ground state, $u \sim r \exp{(-r)}$ satisfies this requirement. The result is that many expectation values will not be finite---for instance, $\langle \hat r^{a} \rangle$ will blow up for $a < -1$.

A related issue is that both constraints, equations~\eqref{eq:reality} and~\eqref{eq:eom} are both derived by assuming self-adjointness---in the former case of $\mathcal O$ and in the latter case, of the Hamiltonian. For example equation~\eqref{eq:eom} is derived with the assumption that $H$ is a self-adjoint operator, meaning that
\begin{align}
    \langle \psi | H \varphi\rangle = \langle H \psi | \varphi\rangle
\end{align}
where $\psi$ and $\varphi$ are wavefunctions. This is true only subject to certain restrictions on the wavefunctions $\psi$ and $\varphi$. We can derive self-adjointness for a restricted family of wavefunctions by integrating by parts:
\begin{align}
    \int_0^\infty dr \, \psi \left( -\frac{1}{2} \partial_r^2 + V \right) \varphi \ &= \  \int_0^\infty dr \, \varphi \left( -\frac{1}{2} \partial_r^2 + V \right)\psi  + \frac{1}{2} (\varphi \partial_r \psi - \psi \partial_r \varphi) \big|^\infty_0\text.
\end{align}
The second boundary term vanishes if the $\psi$ and $\varphi$ are good wave-functions with bounded energy, because bounded energy means that they vanish as $r \to 0$. However it is possible that by multiplying a good wave-function $\psi$ by an operator $\mathcal O$, we will spoil self-adjointness because $\mathcal{O}\psi$ will not vanish fast enough as $r \to 0$. In this case there are two possibilities: (1) the boundary term blows up or (2) they are finite. In case (1) we simply can't use the constraint. In case (2) we may derive constraints that are shifted by a term related to $\psi'(r)|_{r = 0}$. These finite shifts are referred to in \cite{Berenstein:2022ygg} as anomalies. In the case of the Coulomb potential we will see that there is no need to include constraints with anomalies, and indeed there is no advantage to doing so. 

The system can be bootstrapped with the following positive matrices:
\begin{align}
    M_{ij} = \langle \mathcal A_i^\dagger \mathcal A_j \rangle
    \text{ and }
    G_{ij} = \langle \mathcal B_i^\dagger [H ,  \mathcal B_j]  \rangle  \, .
\end{align}
The positivity of $M$ follows directly from the positivity of the Hilbert space norm, while the positivity of $G$ may be derived as the $T \to 0$ limit of the finite temperature constraints derived in~\cite{Fawzi:2023fpg}.

In the case of the Coulomb Hamiltonian, it turns out that we can obtain exact constraints with a very small basis. We choose:
\begin{align}
    \{ \mathcal{A} \} = \{ 1, \hat p, \hat V  \} , \qquad \ \{ \mathcal{B} \} = \{ \hat r, \hat r \hat p \} \, .
\end{align}
After all constraints are taken into account, one finds
\begin{align}
    M = \begin{pmatrix}
        1 & \hat r \hat y & 0 \\
         \hat r \hat y &  \hat y & \frac{1}{2}   y \\
        0 & \frac{1}{2}  \hat y&   \alpha \hat r  \hat y \\
    \end{pmatrix} \text{ and } 2 G = \begin{pmatrix}
        1 & \alpha \\ \alpha & \alpha \hat r \hat y
    \end{pmatrix}
\end{align}
where we have defined $\hat y = \frac{1}{\alpha} V^{(1)}(\hat r) = \hat r^{-2}$. In terms of these variables, and making use of the equation of motion, the energy is $E = \langle \hat H \rangle = - \frac 1 2 \alpha \langle \hat r \hat y\rangle$. 
Now, positivity of $M$ implies that $\langle \hat r \hat y\rangle \le \alpha$, and positivity of $G$ implies $\langle \hat r \hat y \rangle \ge \alpha$. Thus we have that $\langle \hat r \hat y\rangle = \alpha$, and therefore obtain the well-known ground-state energy of the Hydrogen atom:
\begin{align}
    E_0 = - \frac{1}{2} \alpha^2 \text.
\end{align}

\subsection{Cartesian coordinates}

To avoid giving the impression that reduction to the radial Schr\"odinger equation is a necessary component of the quantum-mechanical bootstrap, we now perform the same calculation in three-dimensional coordinates. The material in the previous subsection was not original, having previously been discussed in modern papers on the numerical bootstrap; the material in this subsection is also not original, having previously been discussed in Pauli's 1926 paper on the spectrum of hydrogen~\cite{Pauli1926}.

We will work with the operators $\hat x_i$, their conjugate momenta $\hat p_i$, and the radial coordinate $\hat r = \sqrt{\sum_i \hat x_i^2}$. It is straightforward to confirm that
\begin{equation}
	[\hat r, \hat p_i] = i \frac{\hat x_i}{\hat r}
    \text.
\end{equation}
Then by noting that $0 = [\hat r \hat r^{-1}, \hat p]$, we find
\begin{equation}
	[\hat r^{-1}, \hat p_i] = -i \frac{\hat x_i}{\hat r^3}
	\text.
\end{equation}
Finally we compute
\begin{equation}
	\sum_i \Big[\frac{\hat x_i}{\hat r}, \hat p_i\Big]
	= i \frac{2}{\hat r}
	\text.
\end{equation}
We are now prepared to solve the Hamiltonian (at least in the ground state). Define a vector of operators $\hat a_i$ by
\begin{equation}
	\hat a_i = \hat p_i - i \alpha \frac{\hat x_i}{\hat r}
	\text.
\end{equation}
Note that these operators $a_i$ are not to be interpreted as ladder operators. Although we will decompose the Hamiltonian using these operators below, it is not the case that applying $a^\dagger_i$ to an eigenstate yields an eigenstate. The commutator is
\begin{equation}
[a_i,a^\dagger_j] = \frac {2\alpha}{r} \left(\delta_{ij} - \frac{x_i x_j}{r^2}\right)
    \text.
\end{equation}

The system has an $O(3)$ rotational symmetry under which these operators transform in the fundamental representation. We construct a singlet in the obvious way, and using the commutators derived above we immediately find
\begin{equation}
	\frac 1 2 \sum_i \hat a^\dagger_i \hat a_i
	= \frac 1 2 \sum_i \hat p_i^2 + \frac i 2 \sum_i \Big[\frac{\hat x_i}{\hat r},\hat p_i\Big]
	+ \frac{\alpha^2}{2}
	= \hat H_{\mathrm{Coulomb}} + \frac{\alpha^2}{2}
\end{equation}
Therefore $\langle \hat H_{\mathrm{Coulomb}} \rangle \ge - \frac{\alpha^2}{2}$, which is in fact a tight bound.

This is the (noncommutative) sum-of-squares decomposition of the Coulomb Hamiltonian. Noncommutative sum-of-squares problems are dual~\cite{Lawrence:2021msm} to semidefinite programs (in additional to being, themselves, semidefinite programs), and in fact we can write an explicit bootstrap matrix using these operators to obtain the same result:
\begin{equation}
    M_{ij} = \langle a^\dagger_i a_j \rangle\text.
\end{equation}
In the ground state, all off-diagonal elements of this matrix vanish by rotational symmetry, and the only constraints that are used are the statements $M_{ii} \ge 0$.

Note that in this framework there is no discussion of anomalies. Because the only boundary is the spatial boundary $x \rightarrow \pm \infty$, the Hamiltonian is adjoint over all square-integrable wavefunctions, and there are no anomalies to speak of. This would not be the case if we had adopted the formalism of the radial Schr\"odinger equation, as we do for the remaining sections of this paper.

\section{Yukawa potential}\label{sec:yukawa}

The Yukawa potential arises naturally as an effective theory derived from a three-dimensional relativistic quantum field theory by integrating out a (relatively) heavy boson. For example, the effective nucleon-nucleon force is characterized at long distances by a Yukawa arising from pion exchange~\cite{Wiringa:1994wb}.
The Hamiltonian of a single particle in the Yukawa potential is:
\begin{equation}\label{eq:H-yukawa}
    \hat H_{\mathrm{Yukawa}}
    =\frac{\hat p^2}{2 M} - \frac g {\hat r} e^{-\hat r/\rho}
    \text.
\end{equation}
Here $\rho$ has dimensions of length and is inversely proportional to the mass of the force-mediating particle which has been integrated out. The coupling $g$, defined here so that $g > 0$ indicates an attractive potential, has units of energy times length. This system is characterized by the unique dimensionless quantity $Mg\rho$. For this work we adopt the convention $m = g = 1$, and the only parameter which we vary is the length scale $\rho$.

\begin{figure}
    \centering
	\includegraphics[width=0.7\linewidth]{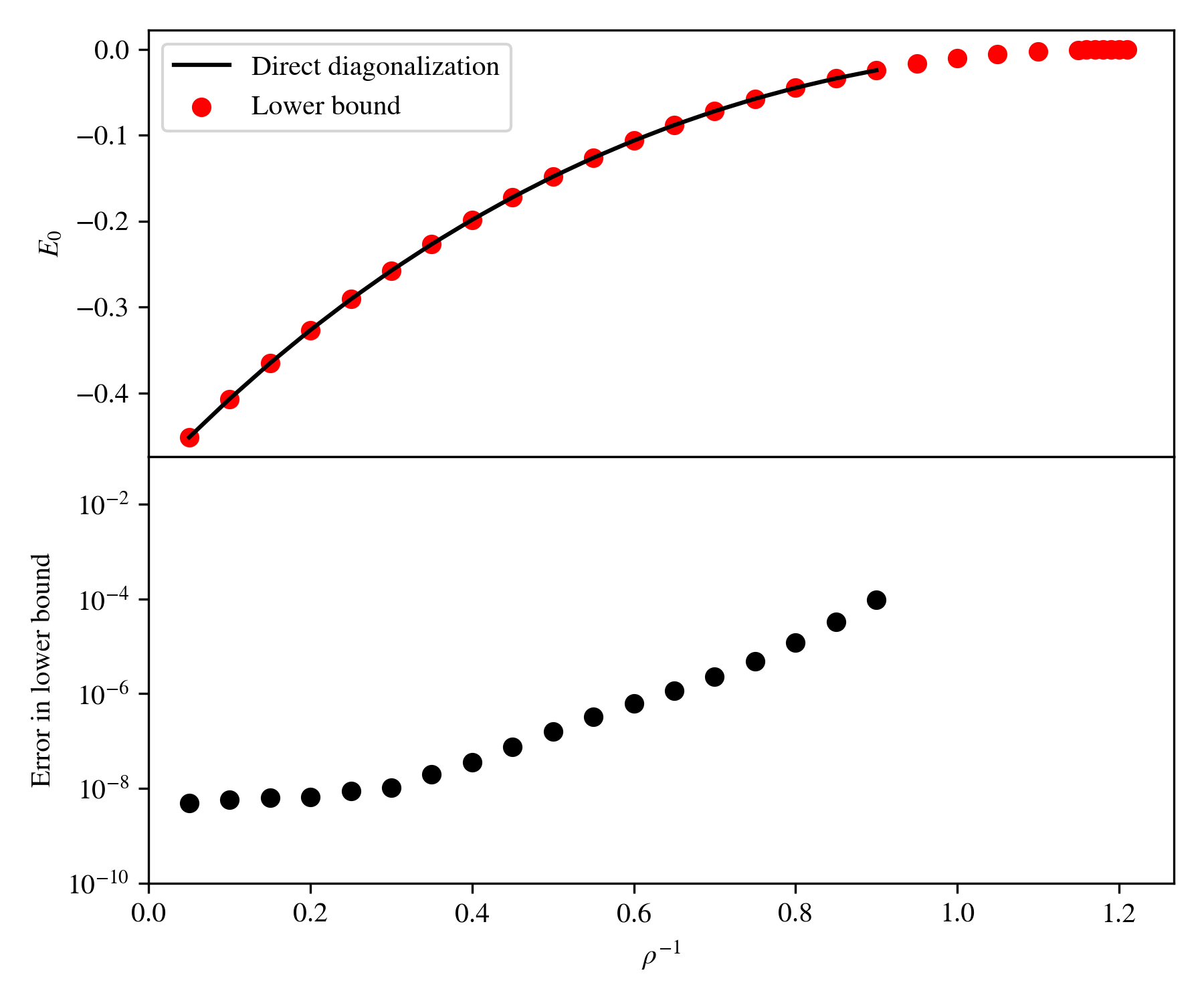}
	\caption{Bootstrap lower bounds on the ground-state energy of the Yukawa potential, across a range of couplings $\rho$. The bottom panel shows the absolute error in these calculations, which is typically better than one part in one million. The comparison points are obtained by direct diagonalization as described in the appendix. Direct diagonalization is trustworthy only below some binding energy; to be conservative, we have truncated the results to the case $E_0 < - 2 \times 10^{-2}$.\label{fig:yukawa}}
\end{figure}

To obtain constraints on the ground state, we define four matrices of the form $M_{ij} = \langle \O^\dagger_i \O_j \rangle$ and two ``ground-state matrices'' of the form $G_{ij} = \langle \O^\dagger_i [H, \O_j ]\rangle$. Each of these matrices is required to be positive semi-definite, and each matrix is defined from a separate basis of operators. The four bases of operators defining matrices $M$ are as follows (eliding hats for readability):
\begin{equation}
	\begin{split}
	\mathcal M^{(1)}&= \{1, 1/r, p, V, rp^2, V p r, pVr, Vp^2r^2, p^2Vr^2, rp,Vr, r^2p^2,Vr^2p, r^2, r, r^2p, Vr^2  \} \\
	\mathcal M^{(2)}&= \{q, q/r, qp, qV, qrp^2, qV p r, qpVr, qVp^2r^2, qp^2Vr^2, qrp,\\
	&\qquad q Vr, qr^2p^2,qVr^2p, qr^2, qr, qr^2p, qVr^2  \} \\
	\mathcal M^{(3)}&= \{ ur^2, ur, ur^2 p, u, u r p, ur^2 p^2 \} \\
	\mathcal M^{(4)}&= \{qur^2, qur, qur^2 p, qu, qu r p, qur^2 p^2, q u /r, q u p, q u r p^2  \}
	\text.
	\end{split}
\end{equation}
Positivity of the corresponding matrices $M$ follows from positivity of the Hilbert-space norm. In addition there are two bases defining matrices $G$:
\begin{equation}
	\begin{split}
	\mathcal G^{(1)}&= \{1, r^3, r^3 p, V r^3, r^2, r, r^2 p, V r^2, rp, Vr, r^2 p^2 \} \\
	\mathcal G^{(2)}&= \{ q r^2, qr, qr^2p, qVr^2, q, qrp, qVr, qr^2p^2, qV, qrp^2 \}\text.
	\end{split}
\end{equation}
Positivity of the corresponding matrices is guaranteed by the fact that we are studying the ground state; see~\cite{Fawzi:2023fpg}. In all expressions above we have used $q = \sqrt{r}$ and $u = \sqrt{-V}$ as convenient denotations.

Together with the affine constraints imposed by commutation relations and the equation of motion $\langle [H,\mathcal O]\rangle = 0$, the above six matrices define a semi-definite program which we solve numerically. The results of the optimization are shown in Figure~\ref{fig:yukawa}, and compared with a numerical solution of the ground state via direct diagonalization. For most couplings, the bootstrap lower bounds obtained with this basis are extraordinarily close to the exact results---usually better than one part in $10^6$. The numerical procedure for obtaining high-precision results by direct diagonalization is somewhat involved; we outline the algorithm in the appendix.

With the basis described above, we do not obtain any non-trivial upper bounds on the ground-state energy. That is, all upper bounds obtained numerically on the ground-state energy are greater than or equal to 0. We leave for future work the question of whether a larger basis is able to provide a non-trivial upper bound.

\subsection{Variables and anomalies}

For this case it will behoove us to be a little more explicit about the calculation. The six bootstrap matrices with the bases defined above lead to a system with 174 independent variables. This is the number of variables \textit{after} using the algebra to choose a canonical order: we put all operators which commute with $\hat p$ to the right of all operators which commute with $\hat x$. For example, $\hat V' \hat p$ is in canonical order, while $\hat p \hat V$ and $\hat p \hat x$ are not.
So there are three types of variables: $r^n p^m$, $r^n V^{(k)} p^m$, and $r^n V^{(k)} V^{(l)} p^m$. 

Next we impose the constraints, which come in three kinds. The first come from requiring that the wavefunctions are real. This, combined with the fact that $\langle O_1 ... O_i \rangle = \langle O_i^\dagger ... O^\dagger_1 \rangle^*$, means that when we reverse the order of operators (which are all Hermitian in our conventions) we get the same thing but with an overall minus sign depending on whether there are an even or odd number of $p$s among the $O_i$.

The second type of constraints come from replacing all derivatives of $V$ with linear combinations of $V$ times powers of $r$. For Yukawa, for instance, we have that
\begin{align}
    V'(r) = -\left(g + \frac{1}{r} \right) V \, , \qquad V''(r) = \left( \frac{2}{r^2} + \frac{2 g}{r} + g^2 \right)V\, ,
\end{align}
and so on. 

The third type of constraints come from the fact that $\langle [\hat H, O] \rangle = 0$. All in all, these constraints drastically reduce the number of independent variables. For the basis given above in the Yukawa theory, the number of variables remaining after the constraints are imposed is 29. Thus our optimization has 29 decision variables.

There is an important subtlety that needs to be taken into account: for the Yukawa interaction (and all central potentials), where there is a boundary in space at $r = 0$, there can be boundary terms that enter into the constraints. These boundary terms arise in both the reversal constraints and the equations of motion, and it is easy to see that they exist by deriving these constraints using integration by parts. Consider for example $
\langle p^3 \rangle$. Normally reversal constraints would require that this is zero, since it is equal to minus itself. However, when we derive it from the wavefunction we find:
\begin{align}
    \langle p^3 \rangle = i\, \int_0^\infty dr  \psi \psi'''
    &=  -i\, \int_0^\infty dr  \psi''' \psi + 2 i \psi \psi''|_0 -  i \psi' \psi'|_0 
    \text.
\end{align}
Note that the boundary terms at infinity die since the wavefunction falls off quickly at infinity. This leads to
\begin{align}
    \langle p^3 \rangle = \frac{i}{2} \left( 2 \psi(0) \psi''(0) - \psi'(0)^2 \right)
\end{align}
Now we need to use some knowledge of the wavefunctions. The first term must vanish because central potentials will only have finite energy if the wavefunction vanishes at the origin, but the second term could in general be non-zero. The way that we handle this is to give this boundary term a name, 
\begin{align}
    A = \frac{1}{2} \psi'(0)^2 \,.
\end{align}
Then we treat $A$ as another undetermined variable that is optimized over in our SDP. This procedure still leads to the solver finding the correct value of the minimum energy. 

Two things are worth noting. First, $A$ shows up in many places, always linearly. For an example of an equation of motion where $A$ appears, consider $\langle [H, p] \rangle$. Without the anomalies, setting this equal to zero would imply $\langle V' \rangle = 0 $. However this clearly cannot be the case because $V'$ is a positive function on $r > 0$. By computing the true Hamiltonian constraint using integration by parts, one finds that the corrected equation is 
\begin{align}
    \langle [H, p] \rangle = i \langle  V' \rangle = i A \, .
\end{align}

The second important point is that this is not the only possible anomaly: higher-derivative boundary terms are also possible. For Yukawa, with the basis given above, the only anomaly to appear is given by $A$, but for the Gaussian potential with the basis given below, one needs to include $B = 1/2 (\psi'(0) \psi'''(0) )$, for instance.

\section{Gaussian potential}\label{sec:gaussian}
We now consider a Gaussian potential. The setup and formalism is largely the same as in the Yukawa case. Even the physical motivation is similar---a Gaussian potential appears in certain fits to the short-range nucleon-nucleon interaction~\cite{Wiringa:1994wb}. However, the Gaussian potential has reasonably different algebraic structure, and we showcase this example here as a way of showing that the bootstrap method is fairly general.

The Hamiltonian we consider is:
\begin{equation}
    \hat H_{\mathrm{Gaussian}}
    =\frac{\hat p^2}{2 M} -
    \frac{b}{(\sqrt{2\pi}R)^3} e^{-\frac{\hat r^2}{2 R^2}}
	\text.
\end{equation}
As before there are three free parameters: the mass $M$, a coupling strength $b$, and a length scale $R$. We have adopted conventions such that the total strength of the potential (and therefore the scattering phase shift at low momentum) is independent of the length scale $R$. Only one dimensionless parameter exists, namely $b M / R$. For the numerical results below we adopt the convention $M,R=1$.

\begin{figure}
	\centering\includegraphics[width=0.7\linewidth]{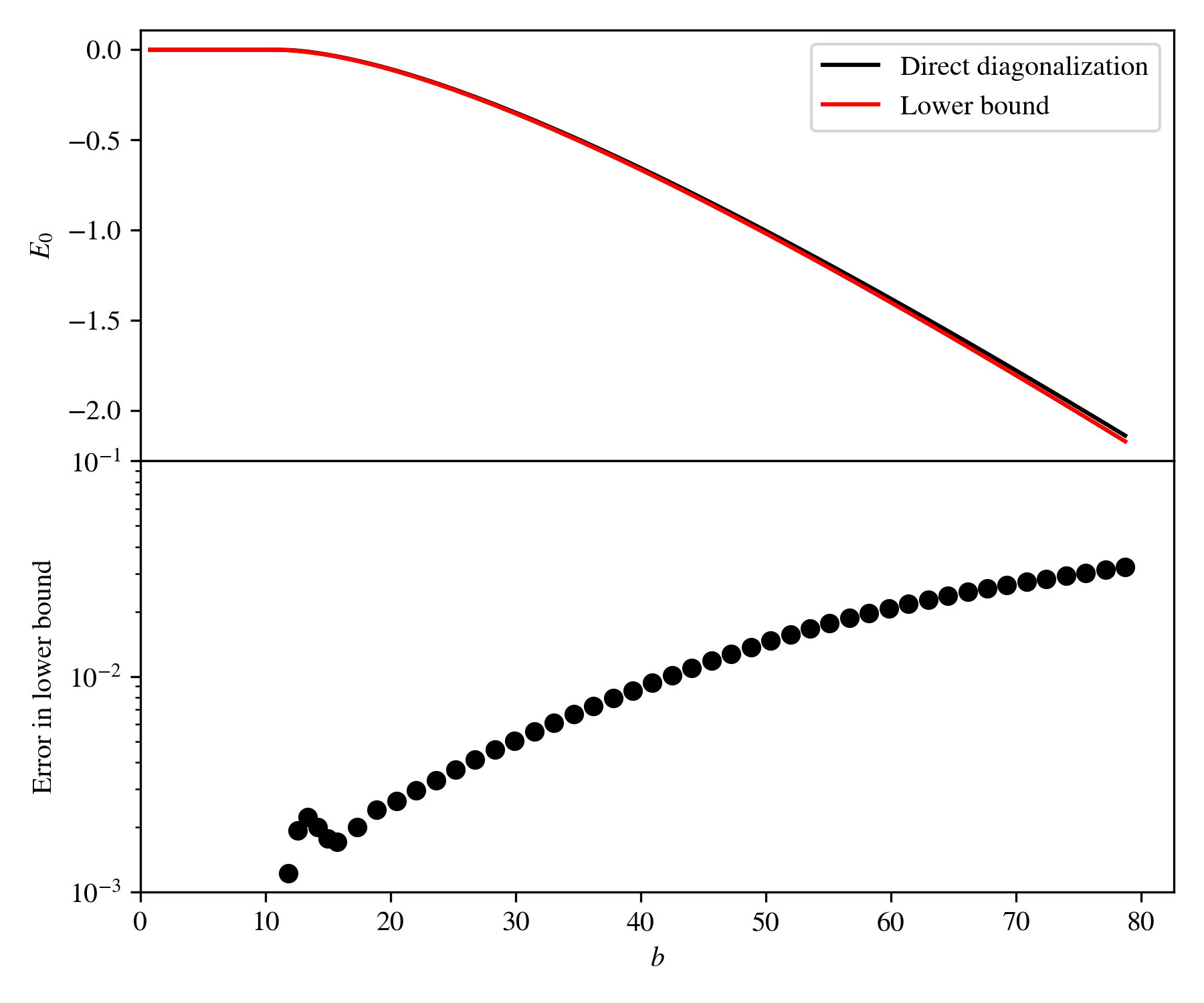}
	\caption{Bootstrap lower bounds on the ground-state energy of the Gaussian potential, as a function of coupling. The relative error is characteristically around 1\%.\label{fig:gaussian}}
\end{figure}

We define six bases of operators. As before, four bases will be used to construct constraints of the form $\langle \mathcal O^\dagger \mathcal O \rangle \ge 0$, and two are used to construct constraints reading $\langle \mathcal O^\dagger [\hat H, \mathcal O]\rangle \ge 0$. The bases corresponding to Hilbert-space positivity (neglecting operator hats for readability) are:
\begin{equation}
	\begin{split}
	\mathcal M^{(1)}&= \{1, 1/r, p, V, rp^2, Vrp, rp,Vr, r^2p^2,Vr^2p, r^2, r, r^2p, Vr^2  \} \\
	\mathcal M^{(2)}&= \{q, q/r, qp, qV, qrp^2,  qVpr,  qrp,\\
	&\qquad qVr, qr^2p^2,qVr^2p, qr^2, qr, qr^2p, qVr^2  \} \\
	\mathcal M^{(3)}&= \{ ur^2, ur, ur^2 p, u, u r p, u/r, u p, ur p^2 \} \\
	\mathcal M^{(4)}&= \{qur^2, qur, qur^2 p, qu, qu r p, q u /r, q u p, q u r p^2  \}
	\text.
	\end{split}
\end{equation}
The two bases which define positivity constraints specific to the ground state are:
\begin{equation}
	\begin{split}
		\mathcal G^{(1)}&= \{ 1/r- i p, rp, r^2 p^2, r, r^2 p, r^2, V, Vr, Vr^2 \}  \\
		\mathcal G^{(2)}&= \{ q, q(1/r-i p), qV, qr p^2, q r p, q V r, qr^2 p^2, qr^2, qr, qr^2p, q V r^2\}
		\text.
	\end{split}
\end{equation}

As in the Yukawa case, these are combined with the affine constraints imposed by commutation relations and the equations of motion to construct a semidefinite program which is solved numerically by SDPB. We do not obtain any non-trivial upper bounds on the ground-state energies, in spite of the use of the $\mathcal G$-constraints. The lower bounds on the ground-state energies, shown in Figure~\ref{fig:gaussian} are qualitatively tight, but much less so than in the case of the Yukawa potential. If these bounds were taken as an estimate of the ground-state energies, the precision would typically only be around $1\%$.


\section{Cornell potential}\label{sec:cornell}
We now turn to the Cornell potential. This is frequently used as a model of the potential experienced by a (heavy) quark-antiquark pair. The Hamiltonian defining the system, in the reduced mass formalism, is
\begin{equation}
    \hat H_{\mathrm{Cornell}}
    =\frac{\hat p^2}{2 M}
    - \frac 4 3 \frac {\alpha_s}{\hat r}
    + \sigma \hat r
    \text.
\end{equation}
Here $M$ is the reduced mass. At short distances the gluon-induced potential is approximated by a Coulomb potential, with coupling $\alpha_s$. At long distances the potential is dominated by the energy of the QCD string connecting the quark-antiquark pair, with string tension $\sigma$.

Note that this potential is qualitatively different from all those considered so far, in that it is not short range, but instead grows without bound at long distances. Moreover, unlike the Yukawa and Gaussian cases, this potential is algebraic: it can be written as a finite sum of rational functions of the coordinate $r$.

Only one combination of the Cornell potential's three couplings is dimensionless, and (up to scalings of energy and distance) this coupling uniquely specifies the spectrum: $\alpha_s$. For the numerical results presented below we work in the convention $\sigma,M=1$.

We follow the same approach as with the Yukawa and Gaussian potentials. Fewer bases, of smaller size, turn out to be sufficient to get tight bounds. The bases we define are:
\begin{equation}
	\begin{split}
	\mathcal M^{(1)}&= \{ 1, 1/r, p, V, rp^2, Vpr, rp, Vr, r^2p^2, Vr^2p, r, r^2p, Vr^2  \} \\
	\mathcal M^{(2)}&= \{ q, q/r, qp, qV, qrp^2, qVpr, qrp, qVr, qr^2p^2, qVr^2p, qr, qr^2p, qVr^2 \}  \\
	\mathcal G^{(1)}&= \{1, rp, Vr, r^2 p^2, r, r^2p, Vr^2, r^3p^2, r^2, r^3p, Vr^3 \}  \\
	\mathcal G^{(2)}&= \{ q, qrp, qVr, qr^2 p^2, qr, qr^2p, qVr^2, qr^2, qr^3p, Vr^3\}\text.
	\end{split}
\end{equation}
\begin{figure}
    \centering
    \includegraphics[width=0.7\linewidth]{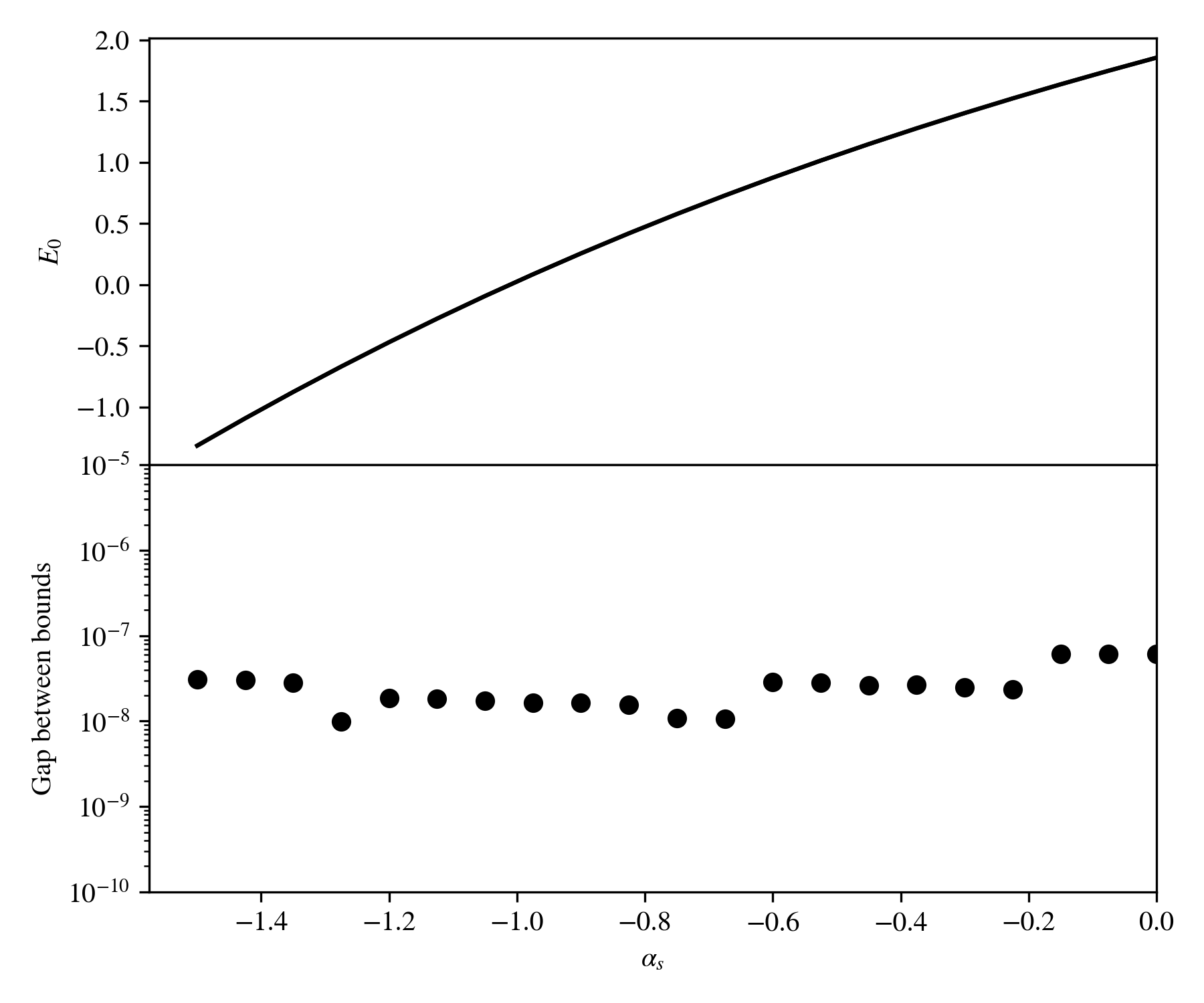}
	\caption{Upper and lower bounds (separated by less than one pixel) on the ground-state energy of the Cornell potential, as a function of the strong coupling $\alpha_s$. For all couplings investigated, the bounds constrain the energy to an interval of size less than $10^{-7}$ (in natural units). These bounds are consistent with numerical results obtained by direct diagonalization (not shown).\label{fig:cornell}}
\end{figure}%
As before, the bases $\mathcal M^{(1)},\mathcal M^{(2)}$ define matrices $M^{(1)},M^{(2)}$ which must be positive semi-definite due to positivity of the Hilbert-space norm. The basis $\mathcal G^{(1)},\mathcal G^{(2)}$ define matrices which are necessarily positive semi-definite only in the ground state of the Hamiltonian. To be concrete, letting $\mathcal O^{(k)}_i$ range over the basis $\mathcal M^{(k)}$, we define
\begin{equation}
	M_{ij} = \langle \big(\mathcal O^{(k)}_i\big)^\dagger \mathcal O^{(k)}_j\rangle
	\succeq 0
	\text.
\end{equation}
Similarly, letting $\tilde{\mathcal O}^{(k)}_i$ range over the basis $\mathcal G^{(k)}$, we define matrices
\begin{equation}
	G_{ij} = \langle \big(\tilde{\mathcal O}^{(k)}_i\big)^\dagger [\hat H, \mathcal O^{(k)}_j]\rangle
	\succeq 0
	\text.
\end{equation}
Now we depart somewhat from the preceding two sections. At each coupling $\alpha_s$ we now define two semidefinite programs. In each we require that all matrices $M,G$ defined above are positive semi-definite, but in one SDP we seek to minimize $\langle \hat H_{\mathrm{Cornell}}$, and in the other we maximize. This results in both a lower and an upper bound on the ground-state energy. Whereas in the Yukawa and Gaussian cases this upper bound was not finite, in this case of the Cornell potential we find that the upper bound is both finite and extraordinarily close to the lower bound.

The numerical results (obtained by SDPB) are shown in Figure~\ref{fig:cornell}. For all couplings, the gap between the bounds in natural units is below one part in $10^7$. For most couplings the same is true for the relative error---only near the critical coupling does the relative error become large.

We can translate these bounds on the energies into bounds on the critical point; that is, the value of $\alpha_s$ at which the ground-state energy vanishes. The critical point was bounded by the following procedure: first, we looked for the smallest possible lower bound that is slightly positive. We found that the lower bound at $\frac 4 3 \alpha_s = -1.34855678$ is equal to $1.95\times10^{-8}$. This means that the critical point must be \textit{below} this value of $\alpha_s$. Then, we find the largest possible value of $\alpha_s$ such that the upper bound is slightly negative. We found that the upper bound on the energy at $\frac 4 3 \alpha_s = -1.34855680$ was equal to $-1.58 \times 10^{-8}$. This means that the critical point must be \textit{above} this point. The result is that the critical point is bounded by
\begin{align}
    -1.34855678 < \frac 4 3 \alpha_s^* <  -1.34855680 \text.
\end{align}

\section{Conformal QM}\label{sec:conformal}

For our final example we consider the following (deceptively simple) Hamiltonian:
\begin{equation}
    \label{eq:Hconf}
    \hat H_{\mathrm{conformal}}
    =\frac{\hat p^2}{2 M} + \frac \lambda{\hat r^2} \, ,
\end{equation}
which is known as ``conformal quantum mechanics'' \cite{Jackiw:1972cb, deAlfaro:1976vlx} because, in $0+1$ dimensions, the coupling $\lambda$ is naively marginal. The nature of the spectrum depends on the coupling, with qualitatively different behavior in three different ranges \cite{Gitman:2009era}:
\begin{align}
    \begin{cases} \frac{3}{8} \leq \lambda & \qquad \text{strongly repulsive} \\
    -\frac{1}{8} \leq \lambda < \frac{3}{8} & \qquad \text{weak medium} \\
    \lambda < -\frac{1}{8} & \qquad \text{strongly attractive} 
    \end{cases}
\end{align}
This theory exhibits rich physics due to the nature of the singularity in the potential. This can lead to violation of probability conservation at $r = 0$ unless care is taken to ensure that the Hamiltonian is Hermitian at $r = 0$---the system is not specified by just a Hamiltonian, but by a Hamiltonian plus a boundary condition for the wave-function at $r = 0$~\cite{Case:1950an}. The strongly attractive phase is particularly interesting. The choice of self-adjoint extension breaks leads to a minimum energy bound state, and then an infinite number of further bound states whose energies all differ by a constant factor which depends on the choice of extension. In other words, conformal symmetry is broken to a discrete subgroup.

In our bootstrap of conformal QM however, we will limit ourselves to studying the full space of permitted density matrices, instead of imposing any restriction to the ground state. Moreover we will not be imposing any particular choice of self-adjoint extension, as we were unable to determine any input to the bootstrap procedure that would accomplish this restriction. Instead, we will bound the relationship between $\langle \hat r^{-1} \rangle$ and the eigenstate energy, in \emph{any} eigenstate of the system. We find that all energies are allowed, as full conformal symmetry requires.

\begin{figure}
    \centering
    \includegraphics[width=0.6\linewidth]{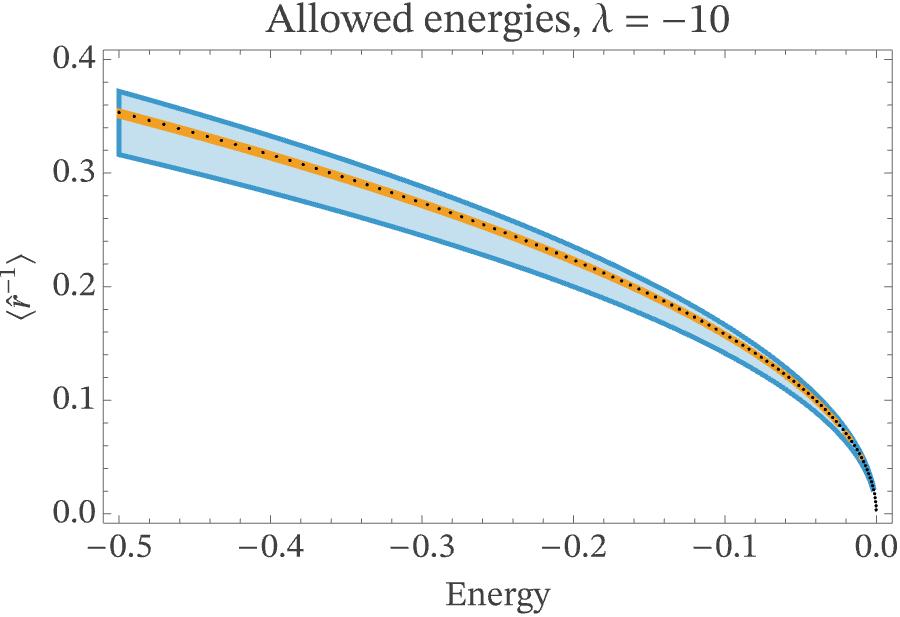}
    \caption{Allowed regions in terms of energies and $\langle \hat r^{-1} \rangle$. The blue region uses two $2 \times 2$ bootstrap matrices, and the orange region uses two $3 \times 3$ matrices. The black dots are the exact results. The energy is unbounded below, but at each $\langle \hat r^{-1}\rangle$ non-trivial bounds are readily obtained.\label{fig:conformal}}
\end{figure}

First note that the positive energy states form a continuum and are delta-function normalized. As such, no $\hat r^n$ has a finite expectation value for positive $n$. It is possible that these could be bootstrapped using a different basis of finite operators; one might then put bounds on expectation values at a given energy. However, for our purposes, it will be more interesting to try to get the discrete states, where we can use the basis $\hat r^n$ and see the values predicted from the above considerations.

First, we have the recursion relation
\begin{align}
	8 (n+1) E \langle \hat r^{t} \rangle + \left( (n+1)n(n-1) - 8 n \lambda \right) \langle  \hat r^{t-2} \rangle = 0 
\end{align}
The recursion relation relates all even powers of $\hat r$ to $\langle 1 \rangle = 1$, and all odd powers to $\langle r^{-1} \rangle$. 
So for a given value of $\alpha$, we have two free parameters, $E$ and $\langle \hat r^{-1} \rangle$. 

We find that the results converge extremely quickly: good results can be obtained with even small bootstrap matrices. Figure~\ref{fig:conformal} was obtained using the matrices
\begin{align}
    M_1 = \begin{pmatrix}
        1 & r \\ r & r^2
    \end{pmatrix} \, , \qquad  M_2 = \begin{pmatrix}
        1/r & 1 \\ 1 & r
    \end{pmatrix} 
\end{align}
for the blue region and 
\begin{align}
    M_1 = \begin{pmatrix}
        1 & r & r^2 \\
        r & r^2 & r^3 \\ 
        r^3 & r^4 & r^5
    \end{pmatrix} \, , \qquad  M_2 = \begin{pmatrix}
        1/r & 1 & r\\ 
        1 & r & r^2 \\
        r & r^2 & r^3
    \end{pmatrix} 
\end{align}
for the orange region. Stronger results can easily be obtained. Here we use $r = \langle \hat r \rangle$ and so on.

\section{Discussion}\label{sec:discussion}

We have investigated the behavior of the quantum-mechanical bootstrap in five systems. The Coulomb potential (which has been treated with this method previously) was reviewed in Section~\ref{sec:coulomb}; as this is an exactly solvable system, it is unsurprising that a small, exact sum-of-squares decomposition of the Hamiltonian is available. The Yukawa and Gaussian potentials were treated in Sections~\ref{sec:yukawa} and \ref{sec:gaussian}, and in both cases we found reasonably tight lower bounds but no non-trivial upper bound. The Cornell potential, central to the phenomenology of heavy mesons, was examined in Section~\ref{sec:cornell}, and tight upper and lower bounds were both obtained. These bounds were sufficient to allow the critical coupling of the Cornell potential to also be bounded, to a precision better than one part in $10^7$. Finally, Section~\ref{sec:conformal} showcased the ability of the quantum-mechanical bootstrap to handle conformal quantum mechanics.

Two (partial) failures of the bootstrap approach should be noted. First, in the cases of the Yukawa and Gaussian potentials, we were unable to obtain any nontrivial upper bounds to the ground-state energy. It is unclear whether any basis (at least of the form we used) would have been sufficient to obtain such upper bounds. Second, the lower bounds on the energy in the Gaussian potential are not particularly tight, having an error of around $1\%$.

Nevertheless, broadly speaking, these results suggest that, at least in the case of the lower bound on the ground-state energy, the bootstrap may be relied upon to converge (in the limit of large bases) to the true ground-state energy. Rigorously establishing this convergence for any family of non-trivial potentials remains an open problem\footnote{In the case of a finite-dimensional Hilbert space the proof is straightforward.}.

\section*{Acknowledgments}

We are grateful to Duff Neill for many conversations regarding the application of the bootstrap to various systems, as well as its likely convergence properties.

S.L.~is supported by a Richard P.~Feynman fellowship from the LANL LDRD program. B.M.~is supported by the Gloria and Joshua Goldberg Fellowship at Syracuse University. Los Alamos National Laboratory is operated by Triad National Security, LLC, for the National Nuclear Security Administration of U.S.~Department of Energy (Contract No.~89233218CNA000001).

\appendix
\addtocontents{toc}{\protect\setcounter{tocdepth}{1}}
\section{Direct diagonalization via Laguerre polynomials}
Obtaining high-precision results suitable for comparison with the quantum-mechanical bootstrap is not completely straightforward, even making use of the radial Schr\"odinger equation. Discretizing Schr\"odinger's equation in the radial coordinate is effective, but often results in numerical errors larger than those obtained by the bootstrap. (The prime symptom of this is that the ground-state energy reported by direct diagonalization can violate the bootstrap-computed bounds.) The purpose of this appendix is to describe the high-precision, but time-efficient, algorithm used in this work to obtain numerical results for comparison in Sections~\ref{sec:yukawa} and \ref{sec:gaussian}.

We assume we have a Hamiltonian $\hat H$ of the form
\begin{equation}
	\hat H = \frac{1}{2M} \nabla^2 + V(\hat r)\text,
\end{equation}
for a potential $V$ which is short-range (i.e.~$V(r)$ decays at least exponentially in $r$). We first obtain a crude initial estimate $E$ of the ground-state energy in this Hamiltonian. This may be done easily by discretizing the Hamiltonian in position space, or by any other reasonable method. The precision of this estimate is not important.

We now define a basis via (generalized) Laguerre polynomials. For each integer $n \ge 0$, we define a wavefunction
\begin{equation}
	\begin{split}
		\tilde\psi_n(r) &= L_n(2\kappa r) r e^{-\kappa r}\\
		&\text{where }
		L_n(x) = \sum_{i=1}^n (-1)^i x^i \frac{(n+2)!}{i! (n-i)! (2+i)!}
		\text.
	\end{split}
\end{equation}
These wavefunctions are orthogonal. Taking $\kappa = \sqrt{-2 M E}$, they are also solutions to the $V=0$ Schr\"odinger equation, and therefore all have (approximately) the correct asymptotic behavior.

The corresponding normalized wavefunctions $\psi_n = \frac{\tilde\psi_n}{|\tilde\psi_n|}$ are easily obtained numerically. We then take $N=16$, and write the Hamiltonian in the basis $\{\psi_0,\ldots,\psi_{N-1}\}$. The resulting ground-state energy is sufficiently accurate to be a meaningful point of comparison for bootstrap results. Critically, this method does not require diagonalizing matrices of any excessive size.

\bibliography{cite.bib}
\bibliographystyle{JHEP.bst}
\end{document}